\documentclass[pra,aps,floatfix,twocolumn,showpacs,reprint,superscriptaddress,linenumbers]{revtex4}

\usepackage[T1]{fontenc}             
\usepackage{mathptmx}                
\usepackage[scaled=.90]{helvet}
\usepackage{graphicx}                
\usepackage{setspace}                
\usepackage[sort&compress]{natbib}   
\usepackage{float}
\usepackage[loose,nice]{units}       
\usepackage{tabularx}                
\usepackage{booktabs}                
\usepackage{amsmath}

\newcommand{\vect}[1]{\ensuremath{\mathbf{#1}}}

\newcolumntype{C}{>{\centering\arraybackslash}X}
\newcolumntype{L}{>{\raggedleft\arraybackslash}X}
\newcolumntype{R}{>{\raggedright\arraybackslash}X}

\begin{document}


\title{One-photon double ionization of helium: a heuristic formula for the cross section}

\author{Morten F{\o}rre}
\email{morten.forre@ift.uib.no}
\affiliation{Department of Physics and Technology, University of Bergen, N-5007 Bergen,
Norway}

\begin{abstract}

Without a formal derivation,
we propose a formula for the total and single-differential
cross in the problem of one-photon double ionization of an atom.
The formula is benchmarked against accurate experimental data for the total cross
section of helium. 
Furthermore, a direct comparison with {\it ab initio} calculations
for the double ionization of Li$^+$  
suggests that the framework is valid for 
the entire helium isoelectronic sequence. 
To this end, we introduce a formula for the double 
ionization of lithium, as well as for the triple ionization of lithium and
beryllium.

\end{abstract}


\maketitle


Double photoionization 
of helium by a single photon has been 
studied for a period of more than 40 years
since the pioneering work of 
Byron and Joachain~\cite{Byron}, who pointed out
the importance of electron correlation in the process.
Such correlated processes pose many challenges, 
and it is only during the last 15 years or so that 
quantitative agreement between 
experiment~\cite{Samson1998} and
theory~\cite{Pont_1995,Pont_1995_L571,Kheifets_1996,Marchalant,Kheifets1998,Rost_1996,Meyer_1997,
Qiu_1998,Nikolopoulos_2001,Hugo_2001,Colgan_2002_2,Foumouo2006,Palacios_2008_2,Nepstad_PRA_2010}
for the total cross section was obtained, and
a rather complete understanding of the breakup process in helium has
emerged~\cite{Briggs,Huetz,Anatoli,Cocke_1998,Malegat_2000,Dawson_2001,Knapp,
Schneider2002,Pattard_Half_collision_Rap}.
A simple analytical formula for the shape of single-photon multiple ionization
cross sections was proposed by Pattard~\cite{Pattard_Shape_Function}.
This shape function contains two parameters,
the position and height of the cross section maximum.

In this work, we propose a formula for the single-differential cross section
in the process of one-photon double ionization of an atom.
The formula contains 
a scaling factor that determines the height of the cross section maximum.
Provided the value of this parameter is set to one, it is shown 
that the formula yields cross sections that are in 
agreement with both theoretical
and experimental double photoionization data for He, Li$^+$ and Li.
Furthermore, it is demonstrated how the framework can be generalized
to account for triple photoionization of lithium and beryllium.

The present work is partly motivated by the idea behind a recent model for  
direct (nonsequential) two-photon
double ionization of helium~\cite{Forre_Super_model_PRL}, i.e., that the
explicit form of the electron-electron interaction is not of crucial importance
in order to obtain a qualitative description of the
electrons' route to the double continuum, as far as the total and single-differential
cross sections are concerned. 
It is here argued  
that the corresponding one-photon double 
ionization event is similarly dictated by the electrons' electric dipole couplings 
to their respective single-particle continua, rather than 
the Coulombic interaction between the electrons. As such, the assumption is that
the electron-electron
interaction merely plays the role of distributing the excess energy 
between the ejected electrons in the excitation process, assuring that the 
total energy of the system is conserved. Keeping in mind that the
electrons are emitted more or less simultaneously (in coincidence) 
in the double ionization process, we simply assume that there is essentially no time 
for the electrons to explore
the explicit geometric form of the repulsive potential at the instant of
ionization, which again suggests that the electron-electron interaction
can be handled in a simple and approximate way. The simplest possible approximate 
model interaction that allows for double ionization by photon impact 
(to first order in perturbation theory), 
controls the energy given to each electron, and which is symmetric with 
respect to both electrons and 
dipole-like for each electron independently, can be written on the 
following heuristic form
\begin{equation}
\label{eq0_1}
H_{int}\propto E(t)z_1z_2.
\end{equation} 
Here $E(t)$ is the electric field modeling the laser pulse, which for simplicity 
is assumed to be
$z$-polarized, and $z_1$ and $z_2$ are the $z$-coordinates of electron $1$ and 
$2$, respectively. 
Although we give no formal proof of the assertion Eq.~(\ref{eq0_1}),
we will nevertheless make use of it in the following to
derive an explicit formula for the single-differential cross section 
in the process of double photoionization of helium and compare it 
with accurate experimental data.

However, before proceeding, we would like
to emphasize that the model interaction in Eq.~(\ref{eq0_1}) is heuristic 
in nature and should be used with caution. For example, 
it fails completely in describing the evolution of the system
in the time after the electrons have been emitted into the continuum, and as 
such it would in general yield incorrect angular distributions.
On the other hand, by construction the interaction 
allows for the possibility that the electrons can absorb the photon
as a unified system, concordant with the model 
of F{\o}rre {\it et al.}~\cite{Forre_Super_model_PRL}.

In the next step of approximation, both electrons are considered to be 
independent particles and the ground (initial) state wave function of the helium
atom is simply approximated by the product ansatz
\begin{equation}
\label{eq1}
\Psi_i(\vect{r}_1,\vect{r}_2)=
\psi_{1s}^{}(\vect{r}_1)\psi_{1s}^{}(\vect{r}_2),
\end{equation}
where $\psi_{1s}^{}$ refers to the ground state of 
the He$^+$-ion.
It should be noted that the overlap between this approximate
wave function and the real ground state wave function of helium is more 
than $90\%$. As such, the assumption is that most of the essential features
relevant for the double ionization process in helium are captured in 
the simplified wave function. 
This is a crucial point in the model presented here. 
Likewise, the final state wave function is approximated by a symmetrized product
of two He$^+$ continuum states,
\begin{equation}
\label{eq2}
\Psi_f(\vect{r}_1,\vect{r}_2)=\frac{1}{\sqrt{N_e!}}\left[
\psi^{}_{E_1}(\vect{r}_1)\psi^{}_{E_2}(\vect{r}_2)+
\psi^{}_{E_2}(\vect{r}_1)\psi^{}_{E_1}(\vect{r}_2)
\right],
\end{equation}
with $N_e=2$ being the number of electrons involved in the ionization process.

Applying lowest order perturbation theory to the resulting system,
with the interaction defined in Eq.~(\ref{eq0_1}), and the initial and final
states defined in Eqs.~(\ref{eq1}) and~(\ref{eq2}), the resulting
single-differential cross section for double photoionization of helium
takes the form
\begin{eqnarray}
\nonumber
\frac{d \sigma}{d E_1}&\propto&
\hbar\omega
\left|\langle \Psi_f\left|z_1z_2\right|\Psi_i\rangle\right|^2\\
\label{eq3}
&=&
\frac{4}{N_e!}\hbar\omega
\left|\langle \psi_{E_1}|z|\psi_{1s}\rangle\right|^2
\left|\langle \psi_{E_2}|z|\psi_{1s}\rangle\right|^2,
\end{eqnarray}
with 
\begin{equation}
\label{eq4}
E_1+E_2=\hbar\omega-E_b,
\end{equation}
$\hbar\omega$ being the photon energy and $E_b=79$ eV the 
binding energy of helium. 
As such, the total binding energy of the system is not
considered a free parameter in the present work.
The coupling
elements in Eq.~(\ref{eq3}) are related to the well known 
one-photon (one-electron) photoionization 
cross section of He$^+$~\cite{Verner_1996} 
via the relation~\cite{Cormier_1995}
\begin{equation}
\label{eq4_b}
\sigma_{He^+} \propto \left(E - E_{1s}\right)  
\left| \langle \psi_{E}|z|\psi_{1s}\rangle \right|^2, 
\end{equation}
where $E_{1s}$ is the energy of the He$^+$ ground state, and
$\sigma_{He^+}$ is the photoionization cross section of He$^+$.

Combining Eqs.~(\ref{eq3}) and~(\ref{eq4_b}),
we propose the following formula for the single-differential cross section
in the process of one-photon double ionization of helium, 
\begin{eqnarray}
\label{eq33}
\frac{d \sigma}{d E_1}=
\frac{C}{N_e!}\cdot\frac{\hbar\omega}{4a_0^2}\cdot
\frac{\sigma_{He^+}^{}(E_1-E_{1s})}{E_1-E_{1s}}\cdot
\frac{\sigma_{He^+}^{}(E_2-E_{1s})}{E_2-E_{1s}}, 
\end{eqnarray}
where $E_1+E_2=\hbar\omega-E_b$, $C$ is a yet unknown 
dimensionless constant, and $a_0$ is the Bohr radius. 
Applying Eq.~(\ref{eq33}), the single-electron
photoionization cross section of He$^+$ is multiplied by two
to account for the statistical weight of having 
two identical electrons in the $1s$ orbital initially. 
Furthermore, the presence of the energy factors 
in the denominators is related to the fact that the photon
is absorbed simultaneously by both electrons via  
a nonresonant transition to the final state of 
each electron~\cite{Forre_Super_model_PRL}. 

\begin{figure}[t]
	\begin{center}
                   \includegraphics[width=8.5cm]{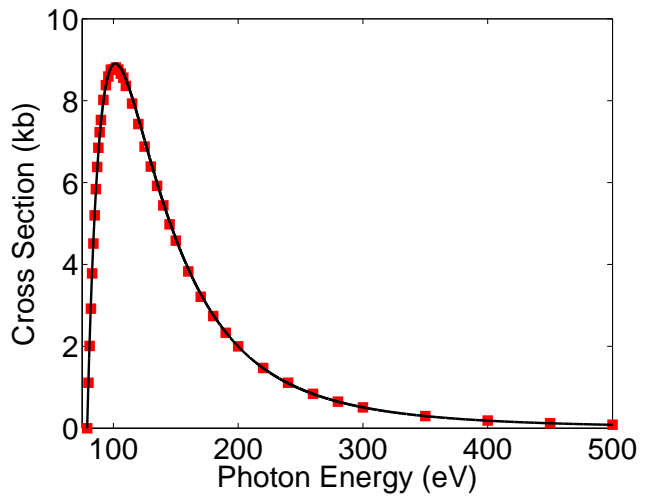}
	\end{center}
	\caption{(Color online) 
Double photoionization cross section of helium versus photon energy.
Black line: model result Eq.~(\ref{eq33}) with $C=1$.
Red squares: experimental results of 
Samson {\it et al.}~\cite{Samson1998}.
} 
	\label{fig1} 
\end{figure}

The one-photon double ionization process of helium 
has been investigated in length in both
theoretical and experimental studies, resulting in close quantitative
agreement in the total cross sections.
In Fig.~\ref{fig1} we compare the total (integrated) cross section
obtained with Eq.~(\ref{eq33}), choosing $C=1$, 
with the accurate experimental data of 
Samson {\it et al.}~\cite{Samson1998}, who stated the accuracy of their
results to be within $\pm 2\%$.  
Provided we choose the value of the (unknown) constant $C=1$,
the model predictions is, within the experimental uncertainty,
in almost exact agreement with the experimental data 
over the entire interval of photon energies considered. 
The value of the parameter $C$ is now determined and will
no longer be considered a free parameter in the rest of this work.
Finally, we would like to note that formula~(\ref{eq33})
also yields single-differential cross sections that are in good
agreement with calculated and measured data for helium.

\begin{figure}[t]
	\begin{center}
                   \includegraphics[width=8.5cm]{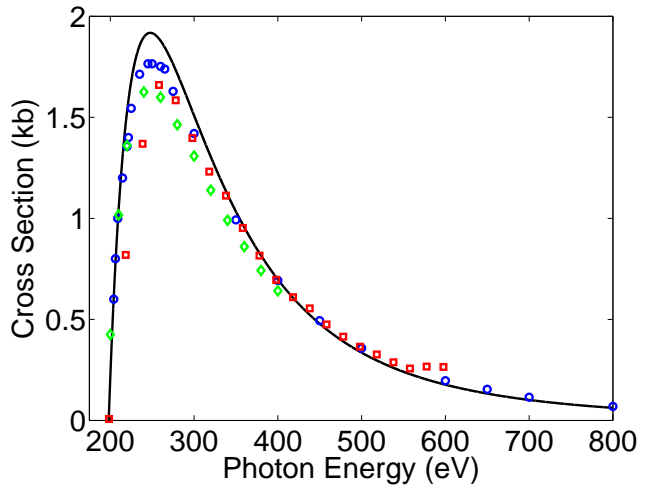}
	\end{center}
	\caption{(Color online) 
Double photoionization cross section of Li$^+$ versus photon energy.
Black line: model result. 
Red squares: theoretical results by van der Hart and Feng~\cite{Hugo_2001}.
Blue circles: theoretical results by Kheifets and Bray~\cite{Kheifets1998}.
Green diamonds: theoretical results by Kleiman {\it et al.}~\cite{Pindzola_2005}.
} 
	\label{fig2} 
\end{figure}

Figure~\ref{fig2} shows the result of formula Eq.~(\ref{eq33})
when applied on the problem of double ionization of Li$^+$, inserting
the corresponding photoionization cross sections of 
Li$^{2+}$  
into the equation.
Theoretical results 
by Kheifets and Bray~\cite{Kheifets1998}, van der Hart and Feng~\cite{Hugo_2001}
and Kleiman {\it et al.}~\cite{Pindzola_2005}
are included for comparison, and
the formula seems to be consistent with the calculated data,
suggesting that it is valid for 
the entire helium isoelectronic sequence.

\begin{figure}[t]
	\begin{center}
                   \includegraphics[width=8.5cm]{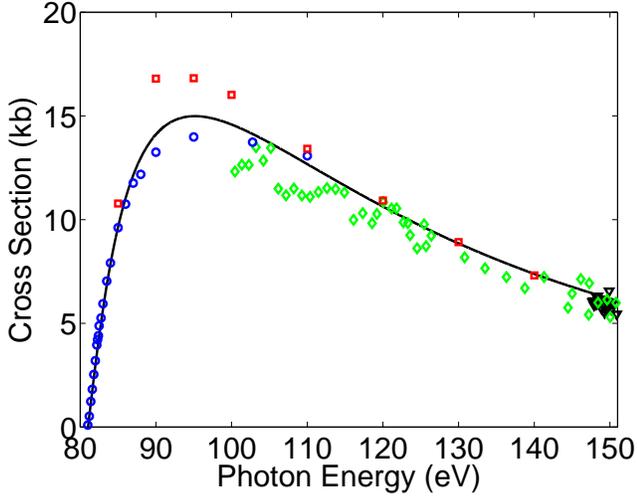}
	\end{center}
	\caption{(Color online) 
Double photoionization cross section of lithium versus photon energy.
Black line: model result Eq.~(\ref{eq333}).
The absolute one-photon photoionization cross 
sections of Li and Li$^+$ are taken from~\cite{Verner_1996}.
Red squares: theoretical results by Colgan {\it et al.}~\cite{PRA_Colgan_2009}.
Blue circles: experimental results by Wehlitz {\it et al.}~\cite{Wehlitz_2002}.
Green diamonds: experimental results by Huang {\it et al.}~\cite{PRA_Huang_1999}.
Black triangles: experimental results by Wehlitz and 
Jurani{\'{c}}~\cite{Wehlitz_PRA_2006_TPDI_Li}.
} 
	\label{fig4} 
\end{figure}

In order to test the validity of the theoretical framework further, 
we now turn to the more challenging case, namely the
double photoionization
of lithium. The assumption is
that it is the outer weakly bound $2s$ electron and one of the tightly bound 
$1s$ electrons that are emitted. This suggests the following formula
for the single-differential cross section,
\begin{eqnarray}
\label{eq333}
\frac{d \sigma}{d E_1}=
\frac{\hbar\omega}{8a_0^2}\cdot
\frac{\sigma_{Li^+}^{}(E_1-E_{1s})}{E_1-E_{1s}}\cdot
\frac{\sigma_{Li}^{}(E_2-E_{2s})}{E_2-E_{2s}}, 
\end{eqnarray}
where $E_{1s}=-75.6$ eV and $E_{2s}=-5.4$ eV are the effective
(single-electron) energies (the negative of the ionization potential) of 
the $1s$ (inner) and $2s$ (outer) electrons, and
$\sigma_{Li^+}$ and $\sigma_{Li}$ are the one-photon 
single-ionization cross sections of Li$^+$ and
Li, respectively~\cite{Verner_1996}.
Figure~\ref{fig4} depicts the 
results for the total cross section for double ionization of lithium,
as obtained by integrating Eq.~(\ref{eq333}), 
and a comparison with previously obtained 
experimental~\cite{PRA_Huang_1999,Wehlitz_2002,Wehlitz_PRA_2006_TPDI_Li} 
and theoretical data~\cite{PRA_Colgan_2009}.
It turns out that the formula yields results that are in good 
agreement with the experimental measurements, 
which is somewhat surprising given the high complexity of the problem.

\begin{figure}[t]
	\begin{center}
		\includegraphics[width=8.5cm]{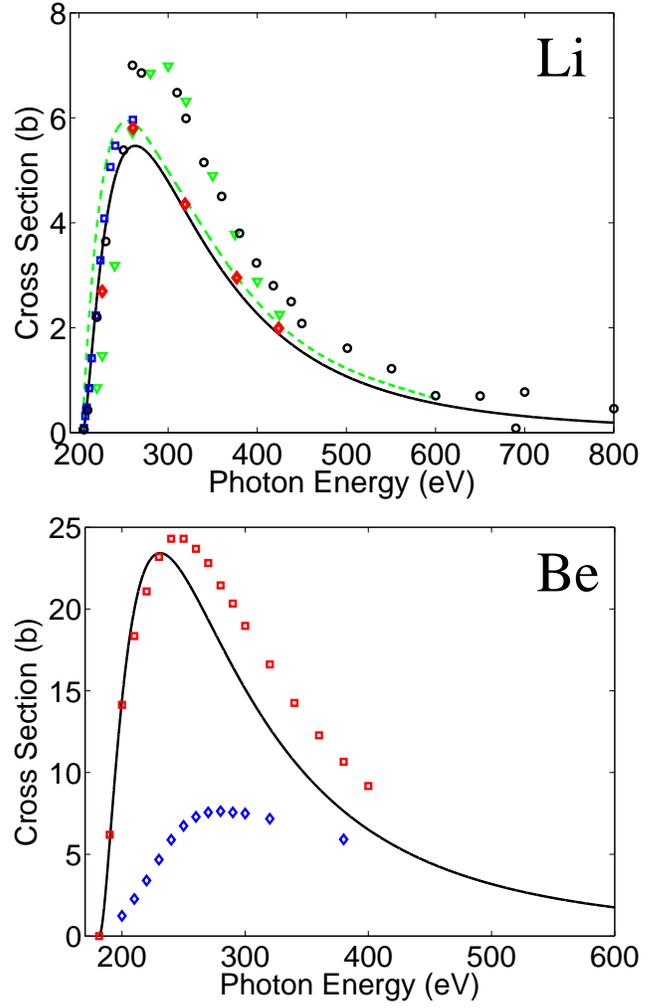}
	\end{center}
	\caption{(Color online) 
Upper panel: Triple photoionization cross section of lithium versus photon energy.
Black line: model result Eq.~(\ref{eqLi}).
Green dashed line: theoretical results by Kheifets and Bray~\cite{Kheifets_Be_2003}. 
Green triangles: theoretical results by Colgan {\it et al.}~\cite{Colgan_PRA_2005}.
Red diamonds, blue squares and black circles:
experimental results by Wehlitz 
{\it et al.}~\cite{Wehlitz_PRL_1998,Wehlitz_PRA_2000,Wehlitz_PRA_2008}.
Lower panel:
Triple photoionization cross section of beryllium versus photon energy.
Black line: model result Eq.~(\ref{eqBe}).
Red squares: theoretical results by Kheifets and Bray~\cite{Kheifets_Be_2003}. 
Blue diamonds: theoretical results by Colgan {\it et al.}~\cite{Colgan_PRA_2005}.
	}
	\label{fig5}
\end{figure} 

Finally, we consider the problem of triple ionization of lithium
and beryllium.
We do this to show
that it is relatively straight 
forward to generalize the framework to consider multiple ionization
processes.
As a matter fact, combining the cross section for
the double ionization of Li$^+$, as obtained in Fig.~\ref{fig2}, 
together with the photoionization cross section of neutral lithium, $\sigma_{Li}$,
according to the rule in Eq.~(\ref{eq33}),
the triple photoionization cross section of lithium is simply given by
\begin{eqnarray}
\label{eqLi}
\frac{d \sigma}{d E_3}=
\frac{\hbar\omega}{24a_0^2}\cdot
\frac{\sigma_{Li^+}^{D}(E_{12}-E_{Li^+})}{E_{12}-E_{Li^+}}\cdot
\frac{\sigma_{Li}^{}(E_3-E_{2s})}{E_3-E_{2s}}. 
\end{eqnarray}
Here $\sigma_{Li^+}^{D}$ denotes the double photoionization cross section 
of Li$^+$ (Fig.~\ref{fig2}),
$E_{Li^+}=-198$ eV is the total energy
of the two bound $1s$ (inner) electrons, $E_{2s}=-5.4$ eV is the energy
of the $2s$ (outer) electron, 
$E_{12}=E_1+E_2$, 
and $E_{12}+E_3=\hbar\omega+E_{Li^+}+E_{2s}$ is the total excess 
energy shared by the electrons in the continuum.  

The result for triple ionization of lithium is shown
in Fig.~\ref{fig5} (upper panel) and compares well with both
experimental~\cite{Wehlitz_PRL_1998,Wehlitz_PRA_2000,Wehlitz_PRA_2008} and 
theoretical~\cite{Kheifets_Be_2003,Colgan_PRA_2005,Emmanouilidou_L99,Emmanouilidou_2006} 
data.
Interestingly, the results are in close agreement with the predictions
of the 
double shake-off model for triple photoionization, proposed
by Kheifets and Bray~\cite{Kheifets_Be_2003}, in particular for the higher
photon energies.

The triple photoionization cross 
section of beryllium is correspondingly given by
\begin{eqnarray}
\label{eqBe}
\frac{d \sigma}{d E_3}=
\frac{\hbar\omega}{24a_0^2}\cdot
\frac{\sigma_{Be}^{D}(E_{12}-E_{Be})}{E_{12}-E_{Be}}\cdot
\frac{\sigma_{Be^{2+}}^{}(E_3-E_{Be^{2+}})}{E_3-E_{Be^{2+}}}, 
\end{eqnarray}
with $\sigma_{Be}^{D}$ being the double photoionization cross section 
of Be (as calculated by Eq.~(\ref{eq33}) inserting the photoionization
cross section of Be$^+$),
$\sigma_{Be^{2+}}$ the one-photon 
single-ionization cross section of Be$^{2+}$,
$E_{Be}=-27.5$ eV the total energy
of the two (outer) bound $2s$ electrons, 
and $E_{Be^{2+}}=-153.9$ eV the 
effective (single-electron) energy 
(the negative of the ionization potential)
of the (active) $1s$ electron.
Furthermore, $E_{12}=E_1+E_2$, 
and $E_{12}+E_3=\hbar\omega+E_{Be}+E_{Be^{2+}}$ is the total excess 
energy shared by the three electrons in the continuum.

The lower panel in Fig.~\ref{fig5} depicts the result
for beryllium together with the theoretical calculations
by Colgan {\it et al.}~\cite{Colgan_PRA_2005}
and  Kheifets and Bray~\cite{Kheifets_Be_2003}.
Quite interestingly, 
the prediction of the formula is again in favor of the
result of the double shake-off model by Kheifets and Bray~\cite{Kheifets_Be_2003},
as far as the total ionization yield is concerned, but
further theoretical and experimental investigations are required
in order to settle the problem definitely.

In conclusion, without a formal derivation, we have proposed a formula 
for the single-differential cross section in the problem of one-photon
double ionization of He, Li$^+$  and Li. 
The corresponding function contains an unknown 
(dimensionless) constant
that dictates the height of the cross section maximum. 
The value of the constant was determined in helium by fitting
with the experimental data of Samson {\it et al.}~\cite{Samson1998}, and
the same value for the constant was used in the other systems. 
Provided a qualified guess for the initial state is taken, the resulting
parametrization is shown to yield results in near quantitative agreement with 
experimental and theoretical data for all considered cases.
Finally, the problem of triple photoionization of lithium and beryllium is studied. 
It is demonstrated that agreement with experimental
and theoretical results can be obtained.
Furthermore, although not shown here, our results are 
consistent with the general shape function proposed 
by Pattard~\cite{Pattard_Shape_Function} and the results
of the half-collision model by Pattard 
and Burgd{\"o}rfer~\cite{Pattard_Half_collision_Rap,Pattard_Half_collision}.

As a final remark, we would like to add that
it is relatively straight forward to generalize the framework to 
account for multiple ionization processes involving more than two or three
electrons, like e.g. the process of
quadruple ionization of beryllium, which is a problem that 
is difficult to pursue within the framework of more 
rigorous {\it ab initio} methods.

%
%
\begin{acknowledgments}
This work was supported by the Bergen Research Foundation (Norway).
The author gratefully acknowledges discussions with 
Peter Lambropoulos, Ladislav Kocbach, S{\o}lve Selst{\o} 
and Aleksander Simonsen, and would like to thank
Ralf Wehlitz, Hugo W. van der Hart, 
Emmanuel Foumouo and Bernard Piraux for sending their data in numerical form.
%
%
\end{acknowledgments}


\end{document}